# PERFORMANCE IMPROVEMENT OF PAPR REDUCTION FOR OFDM SIGNAL IN LTE SYSTEM

Md. Munjure Mowla[1] and S.M. Mahmud Hasan[2]

[1]Department of Electrical & Electronic Engineering, Rajshahi University of Engineering & Technology, Rajshahi, Bangladesh

rimonece@gmail.com

[2] Department of Electronics & Telecommunication Engineering, Rajshahi University of Engineering & Technology, Rajshahi, Bangladesh

etemahmud@gmail.com

## ABSTRACT

*Orthogonal frequency division multiplexing (OFDM) is an emerging research field of wireless communication. It is one of the most proficient multi-carrier transmission techniques widely used today as broadband wired & wireless applications having several attributes such as provides greater immunity to multipath fading & impulse noise, eliminating inter symbol interference (ISI), inter carrier interference (ICI) & the need for equalizers. OFDM signals have a general problem of high peak to average power ratio (PAPR) which is defined as the ratio of the peak power to the average power of the OFDM signal. The drawback of high PAPR is that the dynamic range of the power amplifier (PA) and digital-to-analog converter (DAC). In this paper, an improved scheme of amplitude clipping & filtering method is proposed and implemented which shows the significant improvement in case of PAPR reduction while increasing slight BER compare to an existing method. Also, the comparative studies of different parameters will be covered.*

## KEYWORDS

*Bit Error rate (BER), Complementary Cumulative Distribution Function (CCDF), Long Term Evolution (LTE), Orthogonal Frequency Division Multiplexing (OFDM) and Peak to Average Power Ratio (PAPR).*

## 1. INTRODUCTION

Long term evolution (LTE) is the one of the latest steps toward the 4[th] generation (4G) of radio technologies designed to increase the capacity and speed of mobile telephone networks. It is identical by the third generation partnership project (3GPP) and is an evolution to existing 3G technologies in order to meet projected customer needs over the next decades. LTE has adopted orthogonal frequency division multiplexing (OFDM) for the transmission. OFDM meets the LTE requirement for spectrum flexibility and enables cost-efficient solutions for very wide carriers [1]. The supplementary increasing demand on high data rates in wireless communications systems has arisen in order to carry broadband services. OFDM offers high spectral efficiency, immune to the multipath fading, low inter-symbol interference (ISI), immunity to frequency selective fading and high power efficiency. Today, OFDM is used in many emerging fields like wired Asymmetric Digital Subscriber Line (ADSL), wireless Digital Audio Broadcast (DAB), wireless Digital Video Broadcast - Terrestrial (DVB - T), IEEE 802.11 Wireless Local Area Network (WLAN), IEEE 802.16 Broadband Wireless Access (BWA), Wireless Metropolitan Area Networks (IEEE 802.16d), European ETSI Hiperlan/2 etc [2].

One of the major problems of OFDM is high peak to average power ratio (PAPR) of the transmit signal. If the peak transmit power is limited by either regulatory or application constraints, the effect is to reduce the average power allowed under multicarrier transmission relative to that under constant power modulation techniques. This lessens the range of multicarrier transmission. Furthermore, the transmit power amplifier must be operated in its linear region (i.e., with a large input back-off), where the power conversion is inefficient to avoid spectral growth of the multicarrier signal in the form of intermodulation among subcarriers and out-of-band radiation. This may have a deleterious effect on battery lifetime in mobile applications. As handy devices have a finite battery life it is significant to find ways of reducing the PAPR allowing for a smaller more efficient high power amplifier (HPA), which in turn will mean a longer lasting battery life. In many low-cost applications, the problem of high PAPR may outweigh all the potential benefits of multicarrier transmission systems [3]. A number of promising approaches or processes have been proposed & implemented to reduce PAPR with the expense of increase transmit signal power, bit error rate (BER) & computational complexity and loss of data rate, etc. So, a system trade-off is required. These techniques include Amplitude Clipping and Filtering, Peak Windowing, Peak Cancellation, Peak Reduction Carrier, Envelope Scaling, Decision-Aided Reconstruction (DAR), Coding, Partial Transmit Sequence (PTS), Selective Mapping (SLM), Interleaving, Tone Reservation (TR), Tone Injection (TI), Active Constellation Extension (ACE), Clustered OFDM, Pilot Symbol Assisted Modulation, Nonlinear Companding Transforms etc [4].

## 2. CONCEPTUAL MODEL OF OFDM SYSTEM

OFDM is a special form of multicarrier modulation (MCM) with densely spaced subcarriers with overlapping spectra, thus allowing multiple-access [5]. MCM works on the principle of transmitting data by dividing the stream into several bit streams, each of which has a much lower bit rate and by using these sub-streams to modulate several carriers.

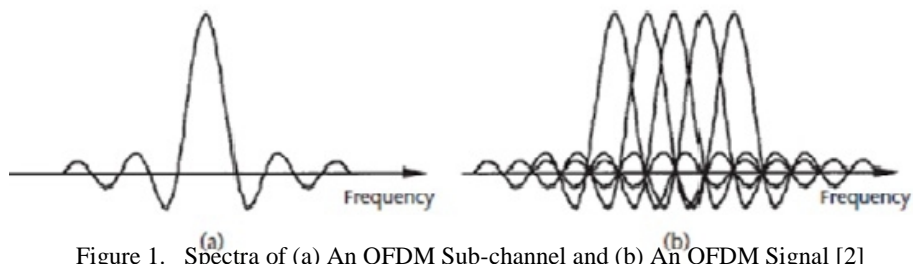

Figure 1. Spectra of (a) An OFDM Sub-channel and (b) An OFDM Signal [2]

In multicarrier transmission, bandwidth divided in many non-overlapping subcarriers but not necessary that all subcarriers are orthogonal to each other as shown in figure 1 (a) [2]. In OFDM the sub-channels overlap each other to a certain extent as can be seen in figure 1 (b), which leads to an efficient use of the total bandwidth. The information sequence is mapped into symbols, which are distributed and sent over the N sub-channels, one symbol per channel. To permit dense packing and still ensure that a minimum of interference between the sub-channels is encountered, the carrier frequencies must be chosen carefully. By using orthogonal carriers, frequency domain can be viewed so as the frequency space between two sub-carriers is given by the distance to the first spectral null [2].

## 2.1. Mathematical Explanation of OFDM Signals

In OFDM systems, a defined number of successive input data samples are modulated first (e.g, PSK or QAM), and then jointly correlated together using inverse fast Fourier transform (IFFT) at the transmitter side. IFFT is used to produce orthogonal data subcarriers. Let, data block of length $N$ is represented by a vector, $X=[X_0, X_{1......}, X_{N-1}]^T$. Duration of any symbol $X_K$ in the set $X$ is $T$ and represents one of the sub-carriers set. As the N sub-carriers chosen to transmit the signal are orthogonal, so we can have, $f_n = n\ f$, where $n\ f = 1/NT$ and $NT$ is the duration of the OFDM data block $X$. The complex data block for the OFDM signal to be transmitted is given by [3],

$$x(t) = \frac{1}{\sqrt{N}} \sum_{n=0}^{N-1} X_n e^{j2\pi n\Delta ft} \qquad\qquad 0 \leq t \leq NT \qquad\qquad (1)$$

Where,

$j = \sqrt{-1}$ , $f$ is the subcarrier spacing and $NT$ denotes the useful data block period. In OFDM the subcarriers are chosen to be orthogonal (i.e., $f = 1/NT$). However, OFDM output symbols typically have large dynamic envelope range due to the superposition process performed at the IFFT stage in the transmitter.

## 3. OVERVIEW OF PAPR

Presence of large number of independently modulated sub-carriers in an OFDM system the peak value of the system can be very high as compared to the average of the whole system. Coherent addition of N signals of same phase produces a peak which is N times the average signal [3]. PAPR is widely used to evaluate this variation of the output envelope. PAPR is an important factor in the design of both high power amplifier (PA) and digital-to-analog (D/A) converter, for generating error-free (minimum errors) transmitted OFDM symbols. So, the ratio of peak power to average power is known as PAPR.

$$PAPR = \frac{Peak\_Power}{Average\_Power}$$

The PAPR of the transmitted signal is defined as [6],

$$PAPR[x(t)] = \frac{\max_{0 \leq t \leq NT} |x(t)|^2}{P_{av}} = \frac{\max_{0 \leq t \leq NT} |x(t)|^2}{\frac{1}{NT} \int_0^{NT} |x(t)|^2\ dt} \qquad\qquad (2)$$

Where, $P_{av}$ is the average power of and it can be computed in the frequency domain because Inverse Fast Fourier Transform (IFFT) is a (scaled) unitary transformation.

To better estimated the PAPR of continuous time OFDM signals, the OFDM signals samples are obtained by $L$ times oversampling [3]. $L$ times oversampled time domain samples are $LN$ point IFFT of the data block with $(L-1)N$ zero-padding. Therefore, the oversampled IFFT output can be expressed as,

$$x[n] = \frac{1}{\sqrt{N}} \sum_{k=0}^{N-1} X_k e^{j2\pi nk/LN} \qquad 0 \leq n \leq NL-1 \qquad\qquad (3)$$

It is known that the PAPR of the continuous-time signal cannot be obtained precisely by the use of Nyquist rate sampling, which corresponds to the case of L = 1. It is shown in that L = 4 can provide sufficiently accurate PAPR results.

The PAPR computed from the L-times oversampled time domain signal samples is given by,

$$PAPR\{x[n]\} = \frac{\max_{0 \le r \le NL-1} |x(n)|^2}{E[|x(n)|^2]} \qquad (4)$$

Where, E{.} is the Expectation Operator.

## 3.1. Motivation of Reducing PAPR

The troubles associated with OFDM, however, are also inherited by OFDMA. Hence, OFDMA also suffers from high peak to average power ratio (PAPR) because it is inherently made up of so many subcarriers [7]. The subcarriers are added constructively to form large peaks. High peak power requires High Power Amplifiers (HPA), A/D and D/A converters. Most radio systems employ the HPA in the transmitter to obtain sufficient transmission power. For the proposed of achieving the maximum output power efficiency, the HPA is usually operated at or near the saturation region [6].

Power efficiency is very necessary in wireless communication as it provides adequate area coverage, saves power consumption and allows small size terminals etc. It is therefore important to aim at a power efficient operation of the non-linear HPA with low back-off values and try to provide possible solutions to the interference problem brought about. Furthermore, the non-linear characteristic of the HPA is very responsive to the variation in signal amplitudes. The variation of OFDM signal amplitudes is very broad with high PAPR. Therefore, HPA will introduce inter-modulation between the different subcarriers and in- traduce additional interference into the systems due to high PAPR of OFDM signals. This additional interference leads to an increase in BER. In order to lessen the signal distortion and keep a low BER, it requires a linear work in its linear amplifier region with a large dynamic range. However, this linear amplifier has poor efficiency and is so expensive.

Therefore, a superior solution is to try to prevent the occurrence of such interference by reducing the PAPR of the transmitted signal with some manipulations of the OFDM signal itself [6]. Large PAPR also demands the DAC with enough dynamic range to accommodate the large peaks of the OFDM signals. Although, a high precision DAC supports high PAPR with a reasonable amount of quantization noise, but it might be very expensive for a given sampling rate of the system. Whereas, a low precision DAC would be cheaper, but its quantization noise will be significant, and as a result it reduces the signal Signal-to-Noise Ratio (SNR) when the dynamic range of DAC is increased to support high PAPR. In addition, OFDM signals show Gaussian distribution for large number of subcarriers, which means the peak signal quite rarely occur and uniform quantization by the ADC is not desirable. If clipped, it will introduce in-band distortion and out-of-band radiation (adjacent channel interference) in to the communication systems. Therefore, the best answer is to reduce the PAPR before OFDM signals are transmitted into nonlinear HPA and DAC.

## 4. CONVENTIONAL CLIPPING AND FILTERING

Amplitude Clipping and Filtering is one of the easiest techniques which may be under taken for PAPR reduction for an OFDM system. A threshold value of the amplitude is fixed in this case to limit the peak envelope of the input signal [8].

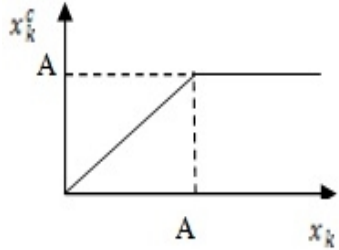

Figure 2. Clipping Function

The clipping ratio (CR) is defined as,

$$CR = \frac{A}{\sigma} \tag{5}$$

Where, A is the amplitude and $\sigma$ is the root mean squared value of the unclipped OFDM signal. The clipping function is performed in digital time domain, before the D/A conversion and the process is described by the following expression,

$$x_k^c = \begin{cases} x_k & |x_k| \leq A \\ Ae^{j\phi(x_k)} & |x_k| > A \end{cases} \qquad 0 \leq k \leq N-1 \tag{6}$$

Where, $x_k^c$ is the clipped signal, $x_k$ is the transmitted signal, A is the amplitude and $\phi(x_k)$ is the phase of the transmitted signal $x_k$.

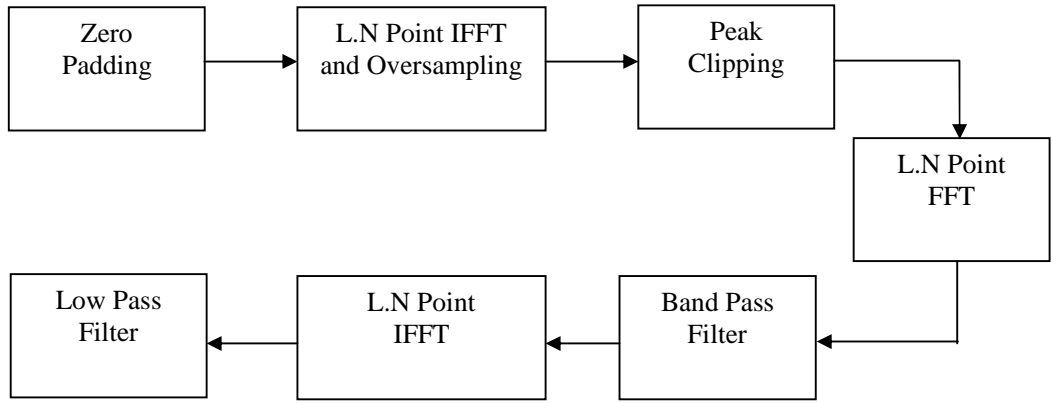

Figure 3. Block Diagram of PAPR reduction using Conventional Clipping & Filtering [9]

In the figure 3, the conventional block diagram of clipping & filtering method is shown which have some limitations described in below.

## 4.1. Limitations of Clipping and Filtering

➢ Clipping causes in-band signal distortion, resulting in BER performance degradation.

➤ Clipping also causes out-of-band radiation, which imposes out-of-band interference signals to adjacent channels. Although the out-of-band signals caused by clipping can be reduced by filtering, it may affect high-frequency components of in-band signal (aliasing) when the clipping is performed with the Nyquist sampling rate in the discrete-time domain[9].

➤ However, if clipping is performed for the sufficiently-oversampled OFDM signals (e.g., L  4) in the discrete-time domain before a low-pass filter (LPF) and the signal passes through a band-pass filter (BPF), the BER performance will be less degraded [10].

➤ Filtering the clipped signal can reduce out-of-band radiation at the cost of peak regrowth. The signal after filtering operation may exceed the clipping level specified for the clipping operation [3].

## 5. PROPOSED CLIPPING AND FILTERING SCHEME

Mentioning the third limitation that is clipped signal passed through the BPF causes less BER degradation, we design a new scheme for clipping & filtering method where clipped signal will pass through the high pass filter (HPF). This proposed scheme is shown in the figure 4. It shows a block diagram of a PAPR reduction scheme using clipping and filtering, where $L$ is the oversampling factor and $N$ is the number of subcarriers. The input of the IFFT block is the interpolated signal introducing $N(L-1)$ zeros (also, known as zero padding) in the middle of the original signal is expressed as,

$$X'[k] = \begin{cases} X[k], & for \quad 0 \le k \le \frac{N}{2} \quad and \quad NL - \frac{N}{2} < k < NL \\ 0 & elsewhere \end{cases} \qquad (7)$$

In this system, the L-times oversampled discrete-time signal is generated as,

$$x'[m] = \frac{1}{\sqrt{LN}} \sum_{k=0}^{LN-1} X'[k] e^{\frac{j2\pi mk}{LN}}, \qquad m = 0,1,...NL-1 \qquad (8)$$

and is then modulated with carrier frequency $f_c$ to yield a passband signal $x^p[m]$.

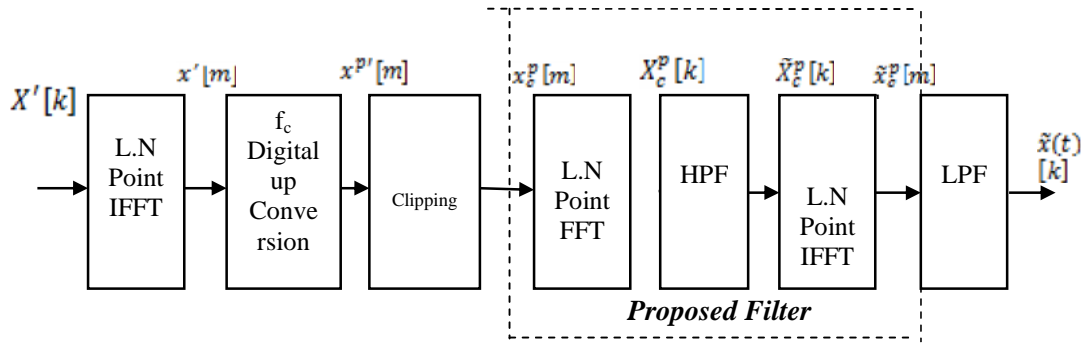

Figure 4. Block Diagram of Proposed Clipping & Filtering Scheme.

Let $x_c^p[m]$ denote the clipped version of $x^p[m]$, which is expressed as,

$$x_c^p[m] = \begin{cases} -A & x^p[m] \leq -A \\ x^p[m] & |x^p[m]| < A \\ A & x^p[m] \geq A \end{cases} \qquad (9)$$

Or,

$$x_c^p[m] = \begin{cases} x^p[m] & if, \ |x^p[m]| < A \\ \frac{x^p[m]}{|x^p[m]|} \cdot A & otherwise \end{cases} \qquad (10)$$

Where, *A* is the pre-specified clipping level. After clipping, the signals are passed through the proposed filter (Composed Filter). The filter itself consists on a set of FFT-IFFT operations where filtering takes place in frequency domain after the FFT function. The FFT function transforms the clipped signal $x_c^p[m]$ to frequency domain yielding $X_c^p[k]$. The information components of $X_c^p[k]$ are passed to a high pass filter (HPF) producing $\tilde{X}_c^p[k]$. This filtered signal is passed to the unchanged condition of IFFT block and the out-of-band radiation that fell in the zeros is set back to zero. The IFFT block of the filter transforms the signal to time domain and thus obtain $\tilde{x}_c^p[m]$.

## 6. DESIGN AND SIMULATION PARAMETERS

In this simulation, a linear-phase FIR filter using the Parks-McClellan algorithm is used in the composed filtering [4]. Existing method [6] uses the band pass filter. But, using this special type of high pass filter in the composed filter, significant improvement is observed in the case of PAPR reduction. The Parks-McClellan algorithm uses the Remez exchange algorithm and Chebyshev approximation theory to design filters with an optimal fit between the desired and actual frequency responses. The filters are optimal in the sense that the maximum error between the desired frequency response and the actual frequency response is minimized. It's a minimax in-band-zero filter which uses three independent equal errors. Table 1 shows the values of parameters used in the QPSK & QAM system for analyzing the performance of clipping and filtering technique [9]. We have simulated the both scheme in the same parameters at first. But, simulation results show the significant improvement occurs in PAPR reduction for proposed method.

Table 1. Parameters Used for Simulation of Clipping and Filtering.

| Parameters | Value |
|---|---|
| Bandwidth ( BW) | 1 MHz |
| Over sampling factor (L) | 8 |
| Sampling frequency, $f_s$ = BW*L | 8 MHz |
| Carrier frequency, $f_c$ | 2 MHz |
| FFT Size / No. of Subscribers (N) | 128 |
| CP / GI size | 32 |
| Modulation Format | QPSK / QAM |
| Clipping Ratio (CR) | 0.8, 1.0, 1.2, 1.4, 1.6 |

## 6.1. Simulation Results for PAPR Reduction

In this paper, simulation is performed for different CR values and compare with an existing method (Yong Soo et.al.) [9]. At first, we simulate the PAPR distribution for CR values =0.8, 1.0, 1.2, 1.4, 1.6 with QPSK modulation and N=128. Then, we simulate with QAM modulation and N=128 and compare different situations.

### 6.1.1 Simulation Results: (QPSK and N=128)

In the existing method, PAPR distribution for different CR value is shown in figure 5 (a). At CCDF $=10^{-1}$, the unclipped signal value is 13.51 dB and others values for different CR are tabulated in the table 2.

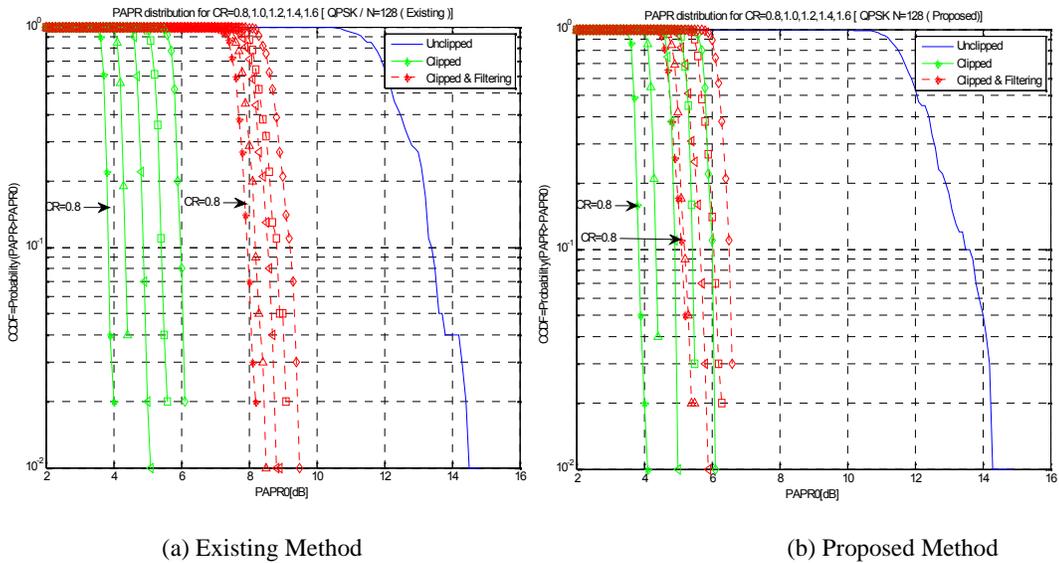

(a) Existing Method         (b) Proposed Method

Figure 5. PAPR Distribution for CR=0.8,1.0,1.2,1.4,1.6  [QPSK and N=128]

In the proposed method, simulation shows the significant reduction of PAPR is shown for different CR values in figure 5(b). At CCDF $=10^{-1}$, the unclipped signal value is 13.52 dB and others values for different CR are tabulated in the table 2. From table 2, we clearly observe that the proposed method reduces PAPR significantly with respect to existing Yong Soo [9] analysis. Proposed method of filter design is done with the same parameters that used in [9].

Table 2. Comparison of Existing with Proposed Method for PAPR value [QPSK and N=128]

| CR value | PAPR value (dB) (Existing) | PAPR value (dB) (Proposed) | Improvement in PAPR value (dB) |
|----------|----------------------------|----------------------------|--------------------------------|
| 0.8 | 7.94 | 5.11 | 2.83 |
| 1.0 | 8.19 | 5.18 | 3.01 |
| 1.2 | 8.55 | 5.65 | 2.90 |
| 1.4 | 8.81 | 6.04 | 2.77 |
| 1.6 | 9.22 | 6.51 | 2.71 |

### 6.1.2 Simulation Results: (QAM and N=128)

The simulation results are now shown for QAM modulation and no. of subscribers, N=128. In the existing method, PAPR distribution for different CR value is shown in figure 6 (a). At CCDF $=10^{-1}$, the unclipped signal value is 13.62 dB and others values for different CR are tabulated in the table 3. In the Proposed method, simulation shows the significant reduction of PAPR is shown for different CR values in figure 6(b). At CCDF $=10^{-1}$, the unclipped signal value is 13.64 dB and others values for different CR are tabulated in the Table 3.

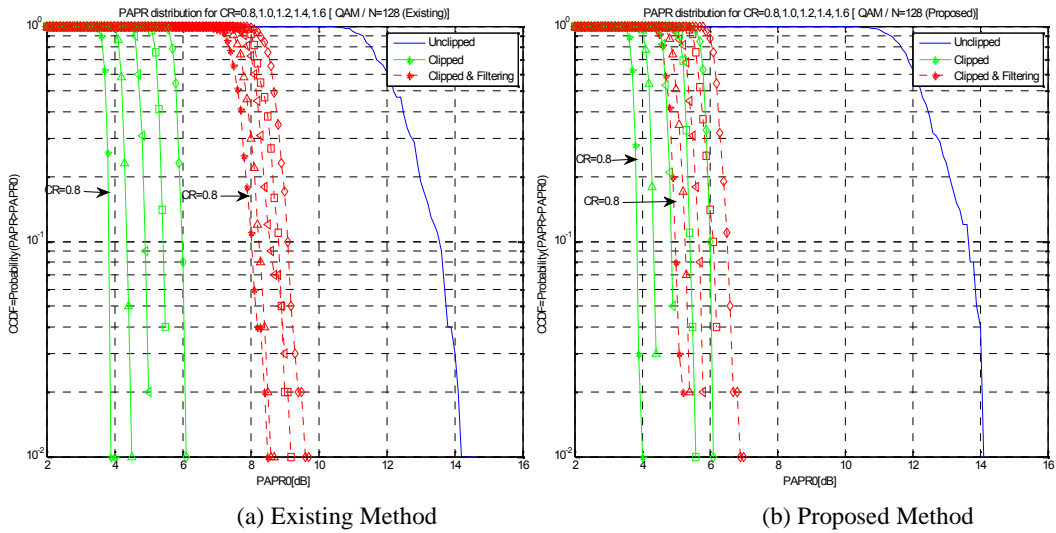

(a) Existing Method          (b) Proposed Method

Figure 6. PAPR Distribution for CR=0.8,1.0,1.2,1.4,1.6 [QAM and N=128]

The existing method and proposed method PAPR distribution values for different CR values are tabulated and differences are shown in table 3. From table 3, we clearly observe that the proposed method reduces PAPR significantly with respect to existing Yong Soo [9] analysis for QAM and N=128 also. So, proposed method works on efficiently for both QPSK & QAM.

Table 3. Comparison of Existing with Proposed Method for PAPR value [QAM and N=128]

| CR value | PAPR value (dB) (Existing) | PAPR value (dB) (Proposed) | Improvement in PAPR value (dB) |
|----------|----------------------------|----------------------------|--------------------------------|
| 0.8 | 8.08 | 4.97 | 3.11 |
| 1.0 | 8.24 | 5.25 | 2.99 |
| 1.2 | 8.56 | 5.67 | 2.89 |
| 1.4 | 8.81 | 6.09 | 2.72 |
| 1.6 | 9.10 | 6.51 | 2.59 |

Now, if we compare the values for different CR values in case of QPSK & QAM to show the effect of modulation on proposed filter design, it is observed that for the same number of subscribers (N=128) & low CR=0.8, QAM provides less PAPR than QPSK. But, at the moderate CR value (1.0, 1.2, 1.4), QPSK results less PAPR than QAM. At the high CR value=1.6, there is no differences between using QAM & QPSK. So, for low CR *(More Amount of Clipping)*, QAM is more suitable than QPSK in case of proposed filter.

## 6.2. Simulation Results for BER Performance

The clipped & filtered signal is passed through the AWGN channel and BER are measured for both existing & proposed methods. Figure 7 shows the BER performance when both clipping and clipped & filtering techniques are used. It can be seen from these figures that the BER performance becomes worse as the CR decreases. That means, for low value of CR, the BER is more.

### 6.2.1 Simulation Results: (QPSK and N=128)

At first, existing method is simulated for QPSK & N=128 for the same data mentioned in table 1 and resulted graph is shown in figure 7(a). Now, simulation is executed for proposed method using same parameters which is shown in figure 7(b) and observed that BER increases with respect to existing method for same value of CR.

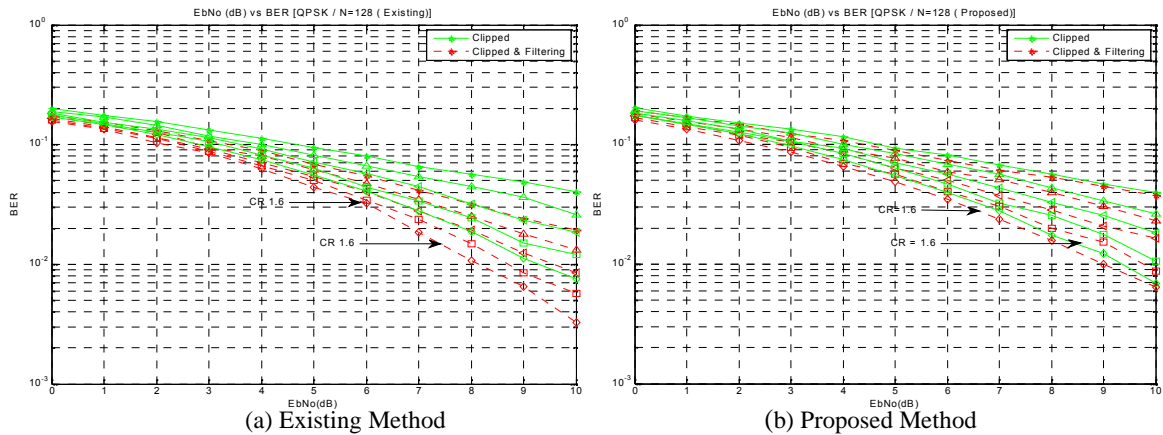

(a) Existing Method         (b) Proposed Method

Figure 7. BER Performance [QPSK and N=128]

From figure 7, it is observed that BER performance is worse in proposed method than existing method for different value of CR. The measured BER value at 6 dB point is tabulated in table 4.

Table 4. Comparison of BER value for Existing & Proposed Method [QPSK and N=128]

| CR value | BER value (Existing) | BER Value (Proposed) | Difference in BER value |
|---|---|---|---|
| 0.8 | 0.0528 | 0.0752 | -0.0224 |
| 1.0 | 0.0451 | 0.0616 | -0.0165 |
| 1.2 | 0.0396 | 0.0492 | -0.0096 |
| 1.4 | 0.0344 | 0.0411 | -0.0067 |
| 1.6 | 0.0289 | 0.0339 | -0.0050 |

### 6.1.2 Simulation Results: (QAM and N=128)

At first, existing method is simulated for QAM & N=128 for the same data mentioned in table 1 and resulted graph is shown in figure 8(a). Now, simulation is executed for proposed method using same parameters which is shown in figure 8(b) and observed that BER increases with respect to existing method for same value of CR.

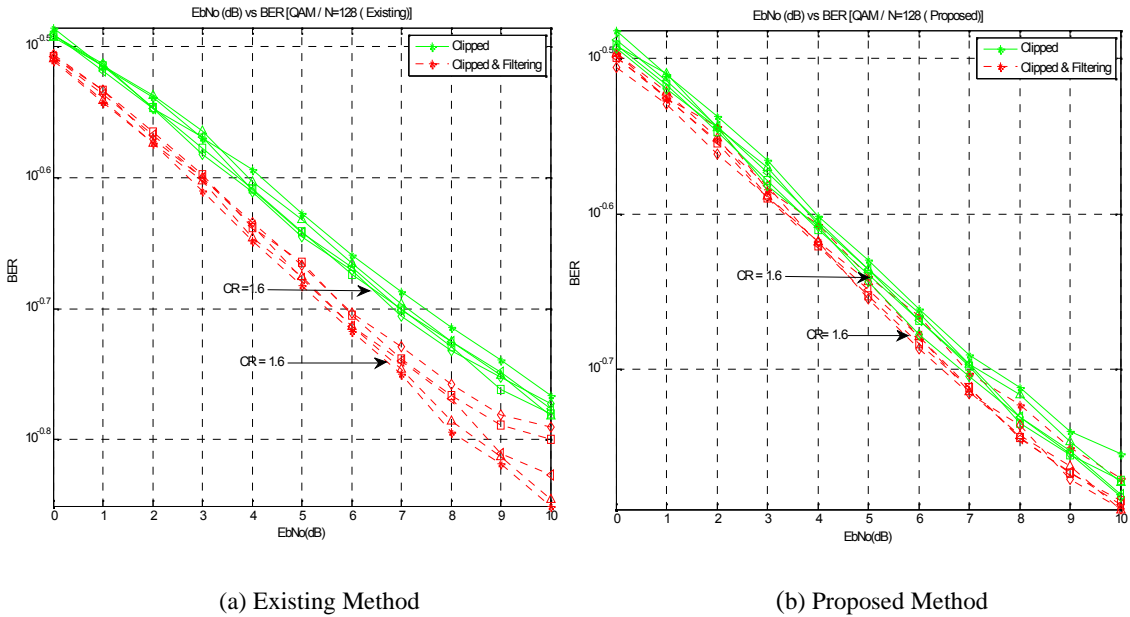

(a) Existing Method          (b) Proposed Method

Figure 8. BER Performance [QAM and N=128]

From figure 8, it is observed that BER performance is worse in proposed method than existing method for different value of CR. The measured BER value at 6 dB point is tabulated in table 5.

Table 5. Comparison of BER value for Existing & Proposed Method [QAM and N=128]

| CR value | BER value (Existing) | BER Value (Proposed) | Difference in BER value |
|---|---|---|---|
| 0.8 | 0.2004 | 0.2138 | -0.0134 |
| 1.0 | 0.1994 | 0.2097 | -0.0103 |
| 1.2 | 0.1951 | 0.2059 | -0.0108 |
| 1.4 | 0.1923 | 0.208 | -0.0157 |
| 1.6 | 0.1906 | 0.2049 | -0.0143 |

BER performance is measured and compared in both the table 4 (QPSK) & table 5(QAM) which shows different CR value for both existing & proposed method in case of same parameter value. For large number of subscribers (N>128), it is found that existing method does not work properly. But, the proposed method works well. From table 4, it is observed that, for CR value (0.8,1.0,1.2,1.4 & 1.6) , the difference magnitude between existing & proposed method are 0.0224,0.0165,0.0096,0.0067 & 0.0050 respectively in QPSK. These BER degradations are acceptable as these are very low values.  From table 5, it is observed that, for CR value (0.8,1.0,1.2,1.4 & 1.6) , the difference magnitude between existing & proposed method are 0.0134,0.0103,0.0108,0.0157 & 0.0143 respectively in QAM. These BER degradations are acceptable as these are very low values.

## 6. CONCLUSION

In this paper, a new scheme of amplitude clipping & filtering based PAPR reduction technique has been analyzed where PAPR reduces significantly compare to an existing method with slightly increase of BER. At first phase, simulation has been executed for existing method with QPSK modulation and number of subscriber (N=128) and then executed for the proposed method for same parameter and observed that PAPR reduces significantly. Next, the simulation has been performed for QAM modulation & N=128 and result shows the considerable improvement in case of QAM also. Then, the proposed method results for both QAM & QPSK modulation with N=128 has been compared. It is observed from the above comparison that for the same number of subscribers (N=128) & low CR=0.8, QAM provides less PAPR than QPSK. But, at the moderate CR value (1.0, 1.2, 1.4), QPSK results less PAPR than QAM. At the high CR value=1.6, there is no differences between using QAM & QPSK. So, for low CR *(More Amount of Clipping)*, QAM is more suitable than QPSK. In the present simulation study, ideal channel characteristics have been assumed. In order to evaluate the OFDM system performance in real world, multipath fading will be a consideration in future. The increase number of subscribers (N) & other parameters can be another assumption for further study.

## Authors

**Md. Munjure Mowla** was born in the Rajshahi, the northern city of Bangladesh. He has completed M.Sc Engineering degree in Electrical & Electronic Engineering (EEE) from Rajshahi University Engi neering & Technology (RUET) in May 2013. He has been working as a Lecturer in Electronics & Telecommunication Engineering department of RUET since November 2010. He obtained B.Sc Engineering degree in Electronics & Communication Engineering (ECE) from Khulna University Engineering & Technology (KUET) in March 2006. He has four years telecom job experience in the operators, vendors, ICX etc of Bangladesh telecom market.  Mr. Mowla has several international & national journals as well as conference papers and three books. He is a member of communication society COMSOC of IEEE, Institutions of Engineers, Bangladesh (IEB) and Bangladesh Electronics Society (BES). His research interest includes advanced wireless communication including green communication, smart grid communication, cognitive radio etc.
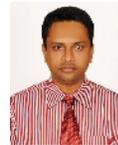

**S. M. Mahmud Hasan** was born in the Rajshahi, the northern city of Bangladesh on 30 January 1991. He has completed B.Sc Engineering degree in Electronics & Telecommunication Engineering (ETE) from Rajshahi University Engineering & Technology (RUET) in September 2012. Now, he is working in the Bangladesh Government Power Development Board (BPDB) as an Engineer. His research interest includes advanced wireless communication (LTE, LTE-A) etc.
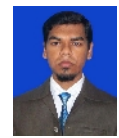